\documentclass[onecolumn]{article}  

\usepackage{natbib}
\usepackage[a4paper, margin=0.6in]{geometry}
\usepackage{amsmath,mathptmx,graphicx,float}

\usepackage{graphicx}
\usepackage{txfonts}
\usepackage{stmaryrd}
\usepackage{xcolor}
\usepackage{authblk}

\newcommand{\ua}{{\bf B}}
\newcommand{\uh}{{\bf u}}
\def\ADD#1{{\textcolor{black}{#1}}}   

\usepackage{fontsize}
\changefontsize[12.75pt]{12pt}

\begin{document} 

\title{The incompressible energy cascade rate in anisotropic solar wind turbulence}

\author[1,2]{N.~Andrés}
\author[3]{F.~Sahraoui}
\author[4]{S.~Huang}
\author[3]{L.Z.~Hadid}
\author[3,5]{S.~Galtier}
        
\affil[1]{\footnotesize Departamento de F\'{\i}sica, Facultad de Ciencias Exactas y Naturales, UBA, Ciudad Universitaria, 1428, Buenos Aires, Argentina}
\affil[2]{\footnotesize Instituto de Astronom\'ia y F\'{\i}sica del Espacio, CONICET-UBA, Ciudad Universitaria, 1428, Buenos Aires, Argentina}
\affil[3]{\footnotesize Laboratoire de Physique des Plasmas, \'Ecole Polytechnique, CNRS, Sorbonne University, Observatoire de Paris, Univ.~Paris-Saclay, F-91128 Palaiseau Cedex, France}
\affil[4]{\footnotesize School of Electronic and Information, Wuhan University, Wuhan, China}
\affil[5]{\footnotesize Institut universitaire de France, France}

\date{}

\maketitle

\begin{abstract}
 {\it Context. }{The presence of a magnetic guide field induces several types of anisotropy in solar wind turbulence. The energy cascade rate between scales in the inertial range depends strongly on the direction of this magnetic guide field, splitting the energy cascade according to the parallel and perpendicular directions with respect to magnetic guide field.}
 
 {\it Aims. }{Using more than two years of Parker Solar Probe (PSP) observations, the isotropy and anisotropy energy cascade rates are investigated. The variance and {normalized fluctuation} ratios, the kinetic and magnetic energies, and  the  normalized cross-helicity and residual energy are studied. The connection between the heliocentric distance, the local temperature of the plasma, and the energy cascade components is made.}
 
 {\it Methods. }{Using exact relations for fully developed {incompressible} magnetohydrodynamic (MHD) turbulence, the incompressible energy cascade rate is computed. In particular, using the isotropy  and 2D and slab  assumptions, the isotropic, perpendicular, and parallel energy cascade rate components are estimated.}
 
 {\it Results. }{The variance anisotropy ratios, for both {velocity} and magnetic fields, do not exhibit a dependence with respect to the heliocentric distance $r$ {between 0.2 and 0.8 au}.  While the {velocity normalized fluctuation} ratio shows a dependence with $r$, the magnetic {normalized fluctuation ratio} does not. A strong correlation between the isotropic and anisotropic energy cascade rates and the temperature is found. A clear dominance of the perpendicular cascades over the parallel cascades as PSP approaches  the  Sun is observed. A dominant 2D cascade and/or geometry over the slab component in slow solar wind turbulence in the largest MHD scales is observed.}
   
\end{abstract}

\section{Introduction}

The solar wind expansion from the Sun is highly non-adiabatic, partly noticed by proton temperatures falling off much more slowly than  is expected for a freely expanding ideal gas \citep[e.g.,][]{P1958,R1995}. Throughout its radial expansion, the solar wind develops a strongly turbulent regime \citep{B2005} that can be characterized by proton density, velocity, temperature, and magnetic field fluctuations \citep{M2011}. Furthermore, large-scale magnetohydrodynamic (MHD) turbulence serves as a reservoir of energy that cascades down to the smallest scales {\citep[e.g.,][]{P1998a,P1998b}} where it can be dissipated by kinetic effects while it heats the plasma {\citep[e.g.,][]{L1998,Sa2009,A2009,H2020a}}. In the MHD inertial range, where the energy is transferred without dissipation through different spatial and temporal scales \citep[e.g.,][]{F1995}, the solar wind exhibits a constant energy cascade rate as a function of such scales \citep{SV2007,Co2015,H2017a,B2020,A2021}, in which the magnetic spectrum presents a -5/3 slope {\citep[e.g.,][]{M1982,L1998,M2021,H2021}}.

The presence of a magnetic guide field ${\bf B}_0$ induces several types of anisotropy in solar wind turbulence on  MHD and kinetic {dissipation} scales \citep[see][]{Ho2012}. In particular, the energy transfer between scales depends strongly on the direction of the magnetic guide field, splitting the energy cascade according to the parallel and the perpendicular directions with respect to ${\bf B}_0$. Several observational results have shown that the solar wind fluctuations at 1 astronomical unit (au) at the largest MHD spatial scales are a combination of field-aligned (or slab) and perpendicular (or 2D) wavevectors  \citep[see][]{Ma1990,Da2005}. \citet{Da2005} used five years of ACE data from near-Earth orbit to investigate the correlation anisotropy of solar wind MHD scale fluctuations and showed that the nature of the anisotropy differs in fast and slow solar winds. In particular, fast winds are  more dominated by fluctuations with wavevectors almost parallel to the local magnetic field, while slow solar winds, which appear to be more fully evolved turbulence, are more dominated by quasi-perpendicular fluctuation wavevectors. \citet{Ad2021} studied anisotropic turbulence in the slow and fast solar wind as a function of the angle between the mean solar wind speed and the mean magnetic field and as a function of the heliocentric distance. Using Solar Orbiter measurements, the authors compared the observed results with the solar wind and with nearly incompressible (NI) MHD turbulence transport model equations \citep{Z1993}, and found agreement between the theoretical and observed results in the slow and fast winds as a function of the heliocentric distance.

Typically, there are two types of fluctuation anisotropy that are recurrently observed in the solar wind, spectral and variance anisotropy \citep[see][]{O2015}. On the one hand, if the components of the fluctuating magnetic (or {velocity}) field have unequal {average} energies, then the field is said to exhibit variance or component anisotropy \citep{Ma2005,W2011}. On the other hand, when the energy distribution at a given spatial ($\ell$) or temporal ($\tau$) scale is not isotropic, the field exhibits  spectral or wavevector anisotropy \citep{M1981,Sh1983,G1995,O2015}. In the present paper we focus our attention on two particular features of anisotropic turbulence, the variance anisotropy ratio and the ratio of fluctuation to mean field for the velocity and the magnetic fields, respectively. The investigation of these anisotropy ratios, the energy cascade rate in the MHD scales, the isotropic and anisotropic models, and their connection with the solar wind temperature are the main objectives of the present paper. 

Using exact relations in fully developed turbulence, it is possible to obtain expressions for the energy cascade rate. Assuming spatial homogeneity and full isotropy, an exact relation for incompressible MHD turbulence can be derived \citep{P1998b,P1998a}. This exact relation provides a precise computation of the amount of energy per unit time and volume $\varepsilon_\text{I}$ (or heating rate) as a function of the velocity and magnetic correlation functions. The MHD exact relation and its connection with the nonlinear energy cascade rate has been numerically and observationally validated for both incompressible and compressible MHD turbulence \citep{WEY2007,M1999,G1997,C2009b,St2009,S2010,B2016c,H2017a,H2017b,A2018b,A2019}, has been generalized to include sub-ion scale effects \citep{A2018,A2019b,H2018,F2019,F2021a}{, and has been extended to include constant velocity shear effects \citep{W2009,W2010}}. Estimations of the isotropic energy cascade rate in the inertial range of solar wind turbulence have been previously computed at 1 au \citep[see][]{M2008,Co2015,B2016c,H2017a} and more recently at small and large heliocentric distances \citep[see][]{B2020,A2021}.

Assuming a 2D and slab cylindrical symmetric geometry, where  the perpendicular cascade rate is considered to depend only on the perpendicular scale and the parallel cascade depends on the parallel direction, \citet{Mac2008} derived a relation for homogeneous incompressible anisotropic MHD turbulence. In particular, they derived expressions for the correlation functions that are applicable to both parallel and perpendicular cascades. Using seven years of solar wind observations from the ACE spacecraft at 1 au, \citet{Mac2008} found a {region with} linear scaling of the energy flux, as is expected for the MHD inertial range. In addition, they found that both fast and slow solar winds exhibit an active energy cascade over an inertial range, with an energy cascade rate in the parallel direction consistently lower than in the perpendicular direction. \citet{St2009} investigated the convergence of third-order structure functions to compute cascade rates in the solar wind using ACE observation at 1 au covering the years from 1998 to 2007. The authors found that a minimum of one year of data is normally required to get good convergence and statistically significant results. They also compared the computed energy cascade rates with previously determined rates of proton heating at 1 au, as determined from the radial gradient of the proton temperature. \citet{S2010} investigated ACE observations of large cross-helicity states using isotropic and anisotropic expression for the energy cascade rate. In contrast to intervals with small helicity values, large helicity states demonstrate a significant back-transfer of energy from small to large scales. 

In the present paper, using a large Parker Solar Probe (PSP) data set (more than 5000 hours in the solar wind), we extend the current state of knowledge of solar wind turbulence in the inner heliosphere by computing the energy cascade rate using both the anisotropic and isotropic relations for fully developed turbulence. Using magnetic field and plasma \ADD{moment} observations between $\sim0.2$ au and $\sim0.8$ au, we investigate how the energy cascade rate is affected not only by the heliocentric distance, but also by the presence of a guide magnetic guide and the consequence anisotropy.

 The study is structured as follows. In Sections \ref{sec:model} and \ref{sec:exact} we present the theoretical incompressible MHD model and a brief description of the anisotropic and isotropic exact relations, respectively. In Section \ref{sec:obs} we briefly describe the PSP observation data set and the conditions that each turbulent event must fulfill. In Section \ref{sec:res} we present the main results of our analysis. Finally, the discussion and conclusions are developed in Section \ref{sec:dis}.

\section{The incompressible MHD model}\label{sec:model}

The three-dimensional (3D) incompressible MHD equations are the momentum equation for the velocity field {\bf u} (in which the Lorentz force is included), the induction equation for the magnetic field {\bf B}, and the solenoid condition for both fields. These equations can be written as
\begin{align}\label{1} 
        &\frac{\partial \textbf{u}}{\partial t} = -\uh\cdot\boldsymbol\nabla\uh  + \ua\cdot\boldsymbol\nabla\ua - \frac{1}{\rho_0}\boldsymbol\nabla(P+P_M) + \textbf{f}_k  + \textbf{d}_k , \\     \label{2} 
    &\frac{\partial \ua}{\partial t} = - \uh\cdot\boldsymbol\nabla\ua + \ua\cdot\boldsymbol\nabla\uh + \textbf{f}_m + \textbf{d}_m , \\  \label{3} 
    &\boldsymbol\nabla\cdot\uh = 0, \\ \label{4} 
    &\boldsymbol\nabla\cdot\ua= 0,
 \end{align}
where the magnetic field is in Alfv\'en velocity units (i.e.,  the real magnetic field is $\textbf{B}\sqrt{4\pi\rho_0}$, where $\rho_0$ is the mean mass density and $\mu$ is the magnetic permeability of the plasma) and $P_M$ is the magnetic pressure. Finally, \textbf{f}$_{k,m}$ are respectively a mechanical and the curl of the electromotive large-scale forcings, and $\textbf{d}_{k,m}$ are respectively the small-scale kinetic and magnetic dissipation terms \citep{A2016b,F2021b}. 

\section{The exact relation in MHD turbulence}\label{sec:exact}

Using Eqs.~\eqref{1}--\eqref{4} and following the usual assumptions for fully developed homogeneous turbulence (i.e., infinite kinetic and magnetic Reynolds numbers and a steady state with a balance between forcing and dissipation) \citep[see, e.g.,][]{A2017b}, an exact relation for incompressible anisotropic MHD turbulence can be obtained as \citep[e.g.,][]{G2018}
\begin{align}\label{exactlaw0}
        -4\varepsilon&= \rho_0\boldsymbol\nabla_\ell\cdot{\bf F},
\end{align}
where {\bf F} is the  incompressible energy flux
\begin{align}\label{flux}
        {\bf F} &= \rho_0\langle (\delta\uh\cdot\delta\uh+\delta\ua\cdot\delta\ua)\delta\uh - (\delta\uh\cdot\delta\ua+\delta\ua\cdot\delta\uh)\delta\ua\rangle,
\end{align}
and $\varepsilon$ is the total energy cascade rate per unit volume. Fields are evaluated at position $\textbf{x}$ or $\textbf{x}'=\textbf{x}+\boldsymbol\ell$; in the latter case a prime is added to the field. The angular bracket $\langle\cdot\rangle$ denotes an ensemble average \citep{Ba1953}, which is taken here as a time average assuming ergodicity. Finally, we  introduced the usual increment definition:  $\delta\alpha\equiv\alpha'-\alpha$. It is worth noting that we do not have access to multi-spacecraft measurements, and therefore it is necessary to assume some sort of symmetry to integrate Eq.~\eqref{exactlaw0} and be able to compute the energy cascade rate $\varepsilon$ \citep[see][]{St2011}. In particular, we  work with two models for the energy cascade rate, an isotropic form for $\varepsilon_\text{I}$ \citep{P1998a,P1998b}, and the anisotropic expressions $\varepsilon_\perp$ and $\varepsilon_\parallel$ respectively for the perpendicular and parallel cascade rates \citep{Mac2008}.

\subsection{The isotropic energy cascade rate}\label{sec:iso}

Assuming the Taylor hypothesis (i.e., $\ell\equiv\tau U_0$, where $U_0$ is the mean plasma flow speed and $\ell=|\boldsymbol\ell|$ is the longitudinal distance) and full isotropy, Eq.~\eqref{exactlaw0} can be integrated and expressed as a function of time lags $\tau$. While Eq.~\eqref{exactlaw0} includes increments in all the spatial directions, the isotropic cascade only includes increments in the longitudinal direction $\ell$ (for single-spacecraft measurements, in the plasma velocity direction $\hat{\bf U}_0$). Therefore, the isotropic energy cascade rate  \citep{P1998a,P1998b} can be {evaluated using only increments in the longitudinal direction}  as
\begin{align}\label{law_iso}
        \varepsilon_\text{I} &= \rho_0\langle [(\delta\uh\cdot\delta\uh+\delta\ua\cdot\delta\ua)\delta{u}_\ell - (\delta\uh\cdot\delta\ua+\delta\ua\cdot\delta\uh)\delta{B}_{\ell}]/(-4\tau U_0/3)\rangle,
\end{align}
where $u_\ell={\bf u}\cdot{\bf \hat U}_0$ and $B_{\ell}={\bf B}\cdot{\bf \hat U}_0$. In particular, the total isotropic energy cascade rate $\varepsilon_\text{I}$ can be expressed as a function of two components: $\varepsilon_1$ proportional to $\delta u_\ell$, and  $\varepsilon_2$ proportional to $\delta B_{\ell}$.

\begin{figure*}
\begin{center}
\includegraphics[width=0.75\textwidth]{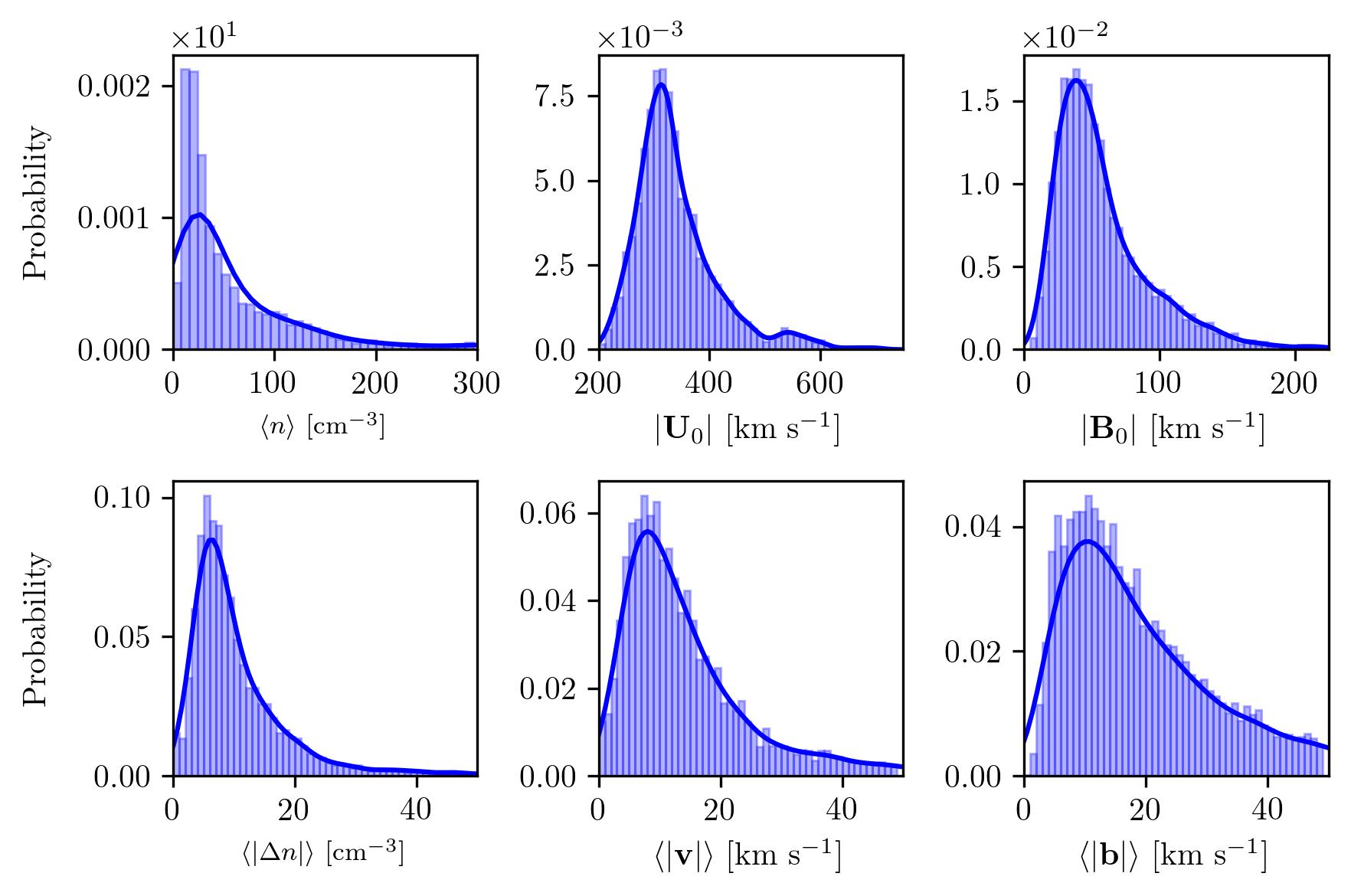}
\end{center}
\caption{Occurrence rates for the proton density, and the proton and Alfv\'en velocity absolute mean values (top) and fluctuations (bottom).}
\label{fig:histo}
\end{figure*}

\begin{figure}
    \centering
    \includegraphics[width=.95\hsize]{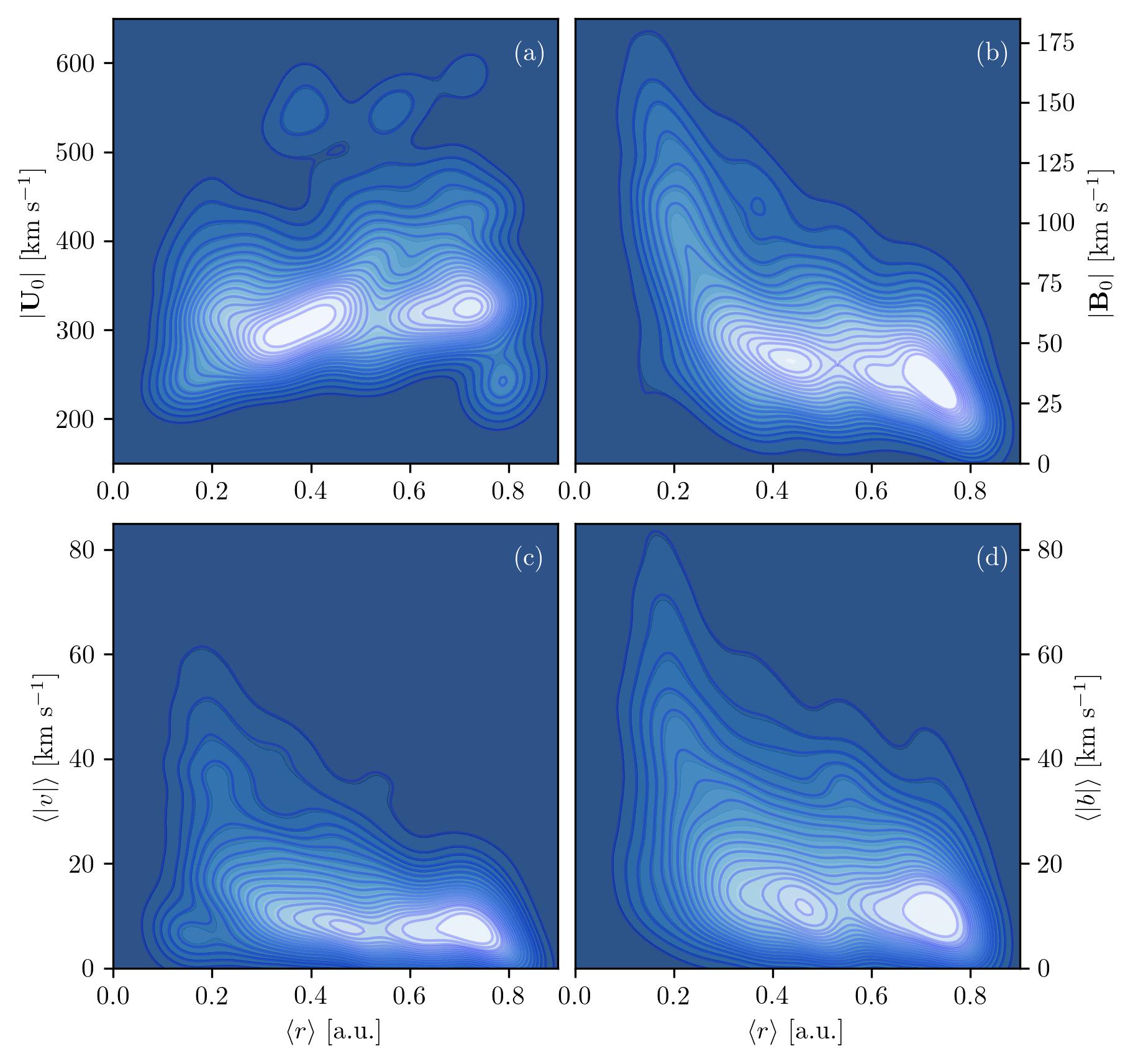}
    \caption{Bivariant KDE for the mean ((a) and (b)) and fluctuating ((c) and (d)) {velocity} and magnetic field absolute values as a function of the heliocentric distance.}\label{fig:kde1}
\end{figure}

\begin{figure}
    \centering
    \includegraphics[width=0.95\hsize]{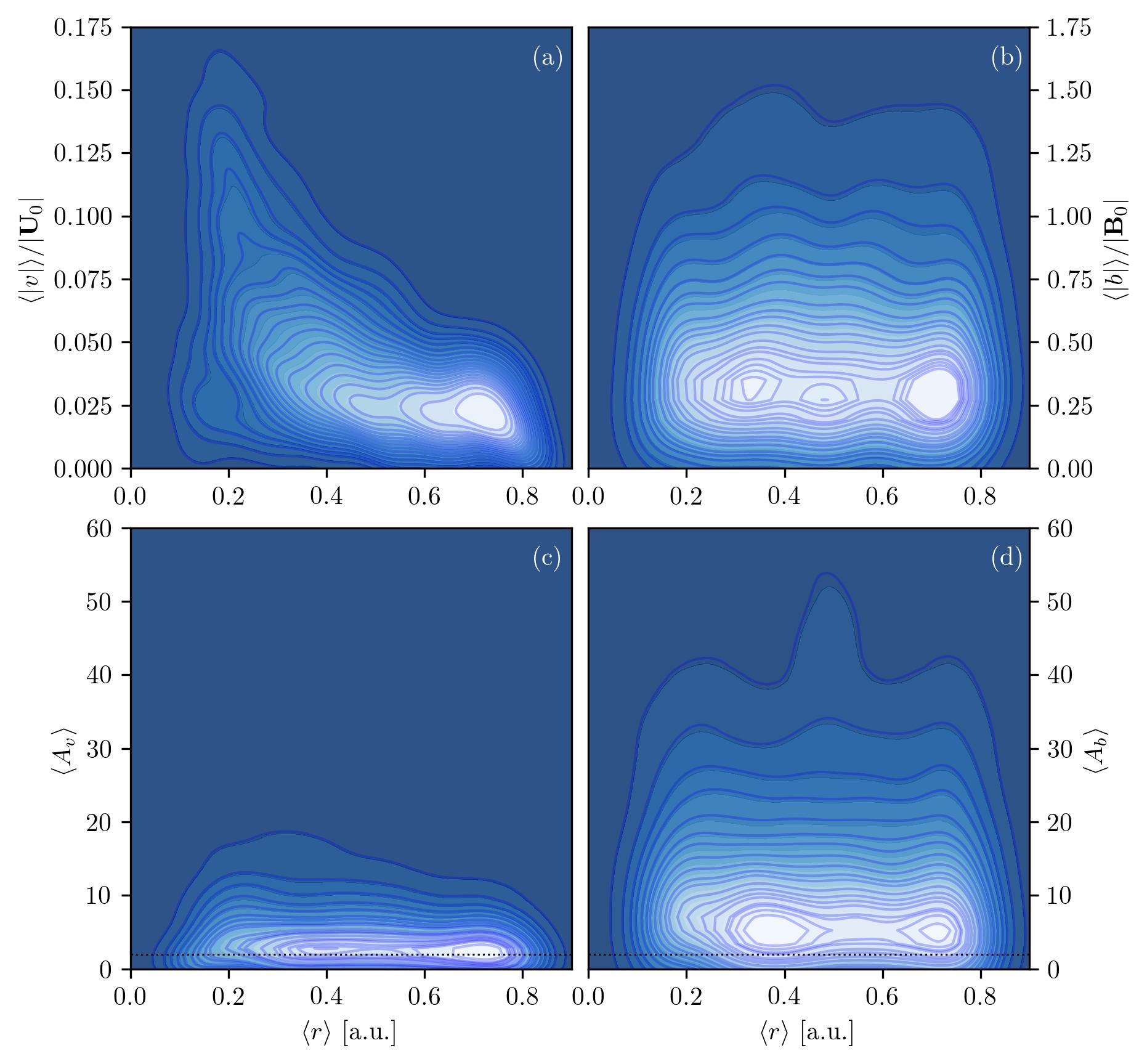}
    \caption{Bivariant KDE for the {normalized fluctuation} ((a) and (b)) and variance anisotropy ((c) and (d)) ratios respectively for the {velocity} and magnetic fields as a function of the heliocentric distance. The dotted lines in (c) and (d) correspond to the isotropic (kinetic or magnetic) energy distribution.}\label{fig:kde2}
\end{figure}

\subsection{The 2D and slab energy cascade rates \citep{Mac2008}}\label{sec:2d1d}

As we discuss in the Introduction, observational results have shown that {an important part of the energy power can be confined to the parallel and perpendicular directions with respect to the magnetic guide field \citep[e.g.,][]{Sh1983,M2004,Da2005,Ou2013}}. Therefore, here we present the hybrid formulation (i.e., 1D plus 2D) that can address the parallel and perpendicular fluctuations, temporal increments, and energy cascade rates \citep[see][]{Mac2008,St2009}. To find expressions for the perpendicular and parallel cascade rates, we use the magnetic field {aligned} basis \citep[e.g.,][]{B1996}, where the velocity and magnetic field observations are properly rotated to leave parallel magnetic fluctuations in one direction. Then, in this particular basis, the ${\hat e}_3$ versor is along the magnetic guide field direction and the unit vectors are
\begin{align}
    {\bf \hat e}_3 &\equiv {\bf \hat e}_{B}, \\
    {\bf \hat e}_2 &\equiv {\bf \hat e}_3\times{\bf \hat e}_1, \\ 
    {\bf \hat e}_1 &\equiv \frac{{\bf \hat e}_U\times{\bf \hat e}_{B}}{|{\bf \hat e}_U\times{\bf \hat e}_B|},
\end{align}
where ${\bf \hat e}_{B} = \langle {\bf B}\rangle / \langle | {\bf B} |\rangle$ and ${\bf \hat e}_U = \langle {\bf u} \rangle / \langle |{\bf u}| \rangle$. Assuming that we have cylindrical symmetry and that the energy flux \eqref{flux} is perpendicular to the mean magnetic field (and depends only on $\ell_\perp$), an expression for the  perpendicular energy cascade rate can be found as
\begin{align}\label{law_perp}
        \varepsilon_\perp &= \rho_0\langle [(\delta\uh\cdot\delta\uh+\delta\ua\cdot\delta\ua)\delta{u}_2 - (\delta\uh\cdot\delta\ua+\delta\ua\cdot\delta\uh)\delta{B}_{2}]/(-2 \tau U_0\sin\theta_{BV})\rangle,
\end{align}
where $u_2={\bf u}\cdot{\bf \hat e}_2$, $B_{2}={\bf B}\cdot{\bf \hat e}_2$ and $\theta_{BV}$ is the angle between ${\bf \hat e}_{B}$ and ${\bf \hat e}_{U}$. On the other hand, still assuming that we have cylindrical symmetry but that the energy flux \eqref{flux} is parallel to the mean magnetic field and depends only on the parallel direction $\ell_\parallel$, an expression for the parallel cascade rate can be found as
\begin{align}\label{law_para}
        \varepsilon_\parallel &= \rho_0\langle [(\delta\uh\cdot\delta\uh+\delta\ua\cdot\delta\ua)\delta{u}_3 - (\delta\uh\cdot\delta\ua+\delta\ua\cdot\delta\uh)\delta{B}_{3}]/(-4 \tau U_0\cos\theta_{BV})\rangle,
\end{align}
where $u_3={\bf u}\cdot{\bf \hat e}_3$ and $B_{3}= {\bf B} \cdot{\bf \hat e}_3$. Finally, the total hybrid energy cascade rate in this model is $\varepsilon_\text{H}=\varepsilon_\perp/2+\varepsilon_\parallel/4$. In the present paper we are interested in computing $\varepsilon_\text{I}$, $\varepsilon_\perp$, and $\varepsilon_\parallel$, which are fully defined by velocity and magnetic field increments that can be estimated from single in situ measurements.

\section{Observations and selection criteria}\label{sec:obs}

We used a data set of PSP observations \citep{Fo2016,K2016,Ba2016,Ka2019,B2019,C2020} covering the period between October 10, 2018, and December 31, 2020. This large data set includes the first six PSP perihelia. We  used the magnetic field and the proton moments from the FIELDS and SPC experiments, respectively. The spurious data (i.e., high artificial peaks) in the SPC moments \citep[see][]{K2016} were removed using a linear interpolation \citep[see][]{B2020,P2020} and the data set was re-sampled to 0.873 s time resolution. In order to analyze the solar wind turbulence on  MHD scales, the data set was divided into a series of samples of equal duration of 60 minutes. This time duration ensures  several correlation times of the turbulent fluctuations at heliocentric distances of less than 1 au \citep[see][]{P2020,H2017a}. As in previous studies \citep[e.g.,][]{A2020,A2021}, we avoided intervals that contained significant disturbances or large-scale gradients (e.g., coronal mass ejection or interplanetary shocks) or rapid flips in the Sun’s magnetic field that reversed direction (i.e., magnetic switchbacks). We further considered only intervals that did not show large fluctuations of the energy cascade rate over the MHD scales; typically, we retained events with $\text{std}(\varepsilon_\text{I})/\text{mean}(|\varepsilon_\text{I}|)<1$ (where std is the standard deviation). 

\section{Results}\label{sec:res}

\subsection{Occurrence rates}

Figure \ref{fig:histo} shows the occurrence rates for the number density, velocity, and magnetic field absolute mean and fluctuation values  for all the events in our data set. In particular, we  separated the velocity and magnetic fields in terms of its mean and fluctuation values as
\begin{align}\label{meanv}
   {\bf u}({\bf x},t) &= {\bf U}_0 + {\bf v}({\bf x},t), \\ \label{meanb}
   {\bf B}({\bf x},t) &= {\bf B}_0 + {\bf b}({\bf x},t), 
\end{align}
where ${\bf U}_0=\langle{\bf u}({\bf x},t)\rangle$, ${\bf B}_0=\langle{\bf B}({\bf x},t)\rangle$ and $\langle\cdots\rangle$ denotes a time averaging operator, which in the present paper is the global mean (i.e., a one hour average). It is worth noting that most of the cases studied in the present paper correspond to slow solar wind (i.e., $|{\bf U}_0| \lesssim 500$ km s$^{-1}$). {Since we want to estimate the incompressible energy cascade rates to ensure the incompressibility approximation, we  keep only the cases where $\langle |\Delta n| / n \rangle < 15 \%$ (where $\Delta n \equiv n - \langle n \rangle$). In other words, we use the full velocity fields in the incompressible MHD exact relation in those events where the velocity fluctuations have only weak compressible effects. However,  we are estimating the incompressible energy cascade rate since velocity fluctuations may still contain a small compressible component.} This leaves us with a data set of $\sim$ 5200 events of one-hour duration each.

Figure \ref{fig:kde1} shows the bivariant kernel density estimation (KDE) for the mean  and fluctuating  velocity and the magnetic fields as a function of the heliocentric distance. A bivariant KDE produces a continuous probability density surface in two dimensions \citep[see][]{W2021}, where brighter regions correspond to regions with more analyzed events. It is worth noting that while the mean velocity field values do not present a statistical dependence with the heliocentric distance, the magnetic guide field and both magnetic and {velocity} fluctuation values strongly decrease as we move away from the Sun. In particular, as we approach  the Sun, the magnetic and kinetic fluctuation levels increase  to the same order ($\sim$ 50 -- 70 km s$^{-1}$). We  return to this point in Section \ref{sec:epsi} when we analyze the isotropic cascade rate.

\begin{figure}
    \centering
    \includegraphics[width=.45\hsize]{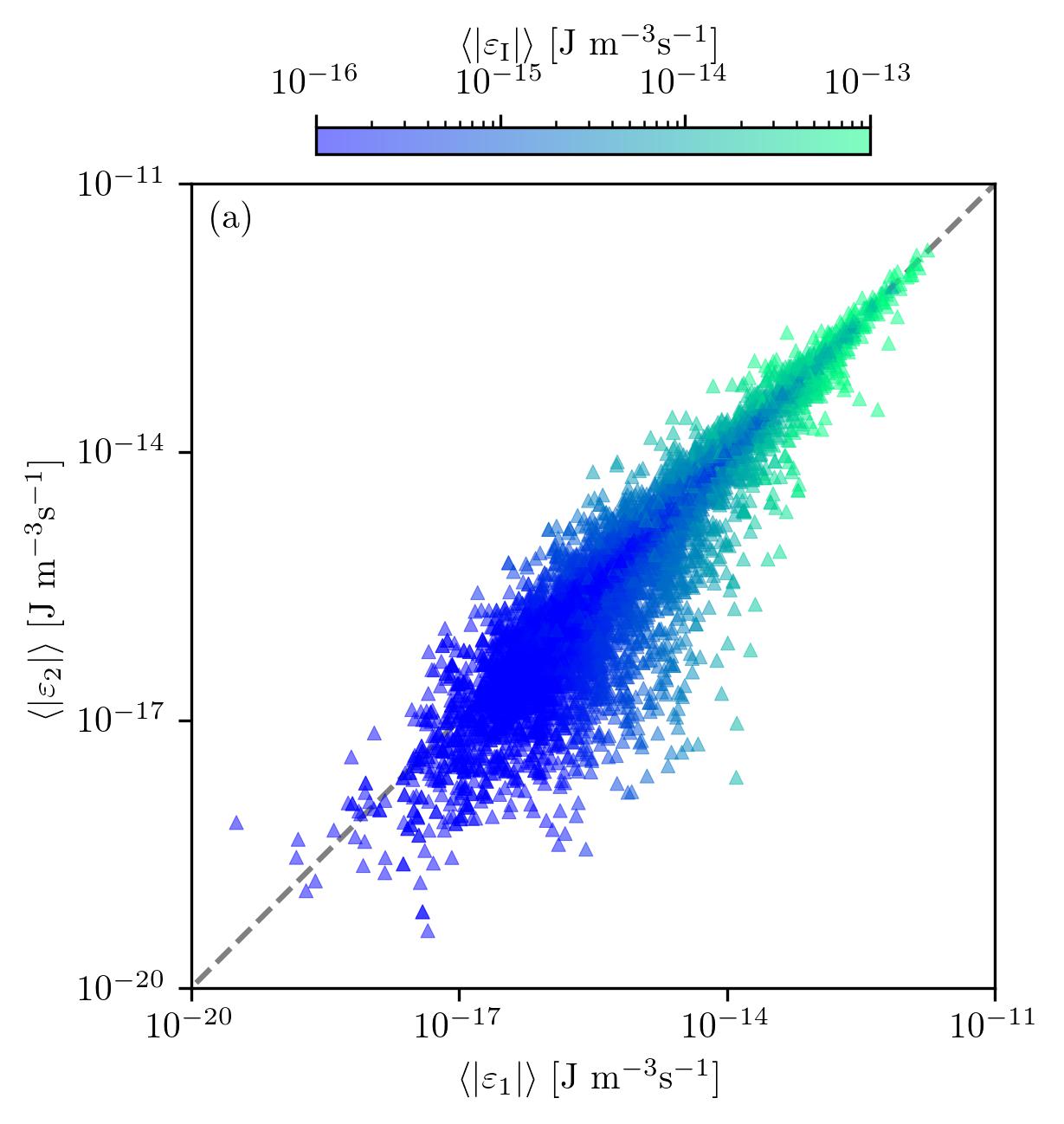}\includegraphics[width=.45\hsize]{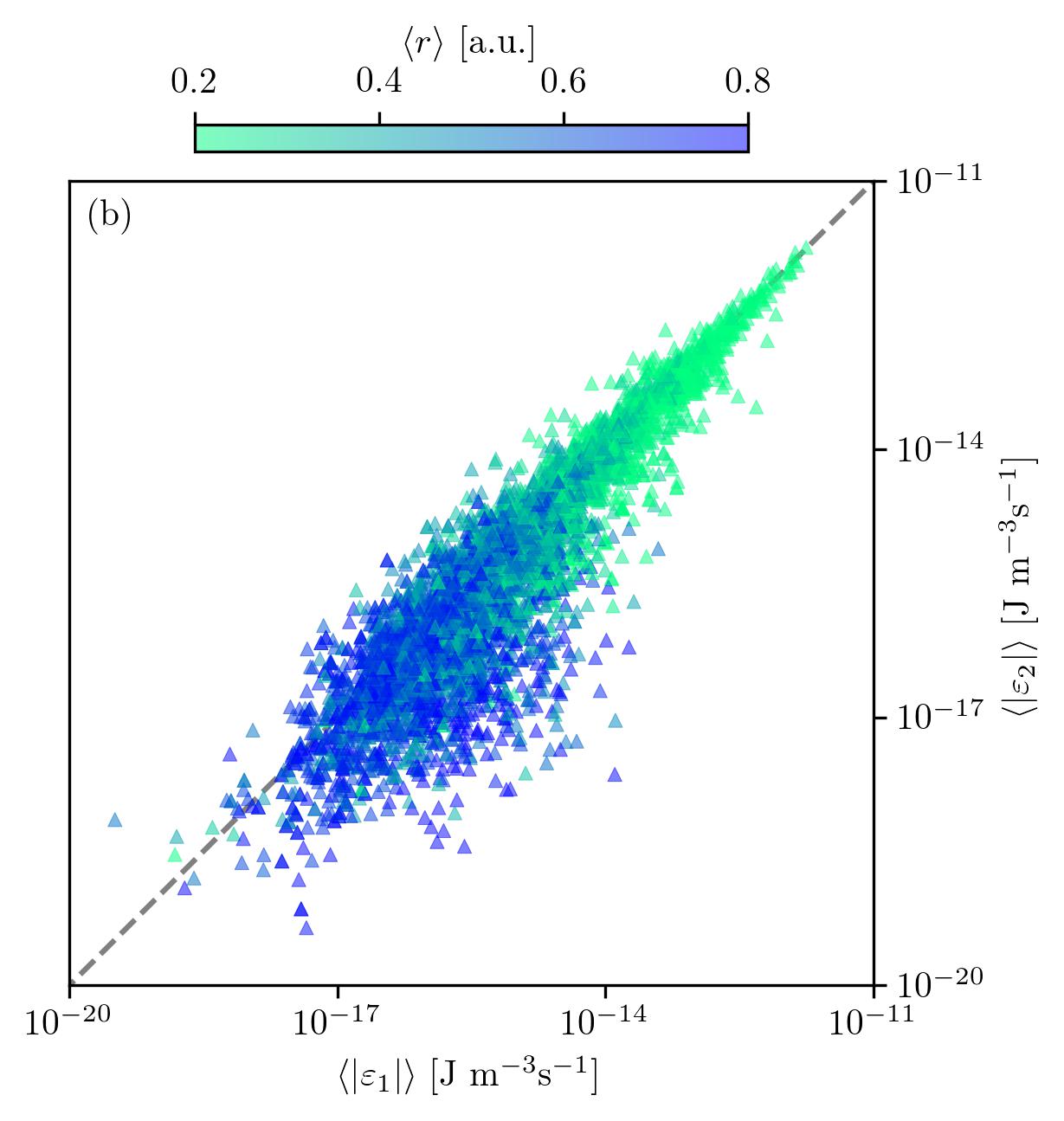}
    \caption{Cascade rate component $\langle|\varepsilon_2|\rangle$ as a function of the component $\langle|\varepsilon_1|\rangle$. In panel (a) the color bar is the total cascade $\langle|\varepsilon_\text{I}|\rangle,$ and  in panel (b) it is the heliocentric distance $\langle r \rangle$.}\label{fig:epsi}
\end{figure}

\subsection{Variance anisotropy and normalized fluctuation ratios}

As we discussed in the Introduction, there are two types of fluctuation anisotropy that are typically observed in the solar wind: spectral anisotropy and variance anisotropy. To quantify them we consider the velocity and magnetic fields in terms of mean values plus fluctuations around these means (see Eqs.~\eqref{meanv} and \eqref{meanb}). On the one hand, if the components of the field have unequal energies (e.g., in {Cartesian} coordinates, departures from $\langle b_x^2 \rangle = \langle b_y^2\rangle= \langle b_z^2\rangle$ for the magnetic field), the field exhibit variance anisotropy {\citep[e.g.,][]{B1971,T2012}}. To quantify this variance anisotropy, we consider the {velocity} and magnetic anisotropy ratios \citep[see][]{O2015} as
\begin{align}
    A_{v} &= \frac{v_\perp^2}{v_\parallel^2},\\
    A_{b} &= \frac{b_\perp^2}{b_\parallel^2},  
\end{align}
where we  employ the magnetic field coordinate system \citep[see][]{B1996}. Variance anisotropy is {scale \citep[e.g.,][]{M2012} and plasma $\beta$ dependent \citep[e.g.,][]{Ou2016}}; however, in the present paper we focus our attention on their values for the largest MHD scales (i.e., one hour mean values). On the other hand, generally speaking, when the energy distribution at a given time scale $\tau$ is not isotropic, we speak of spectral anisotropy. In particular, spectral anisotropy is usually associated with energy cascades that are also anisotropic \citep{O2015,Ho2012}. Moreover, for incompressible MHD turbulence, numerical and observational evidence shows that strong (or even moderate) mean magnetic fields give rise to a suppression of the energy cascade in the parallel direction, and  the perpendicular energy cascade is thus much stronger than the parallel cascade {\citep{Sh1983,O1994,C2000,M2001,Mac2008,St2009,Ou2011,O2013,M2012,A2018b}}. {Therefore, in the present paper we consider the ratio of the fluctuation fields and the mean  as   indicative of spectral anisotropy at   MHD scales for both $\bf u$ and $\bf B$; in other words,  the ratios $\langle |v| \rangle / |{\bf U}_0|$ and $\langle |b| \rangle / |{\bf B}_0|$ are the normalized fluctuation ratios for the {velocity} and magnetic fields, respectively.}

Figure \ref{fig:kde2} show the bivariant KDE for {the normalized fluctuation ratios}  and variance anisotropy ratios  respectively for the {velocity} and magnetic fields as a function of the heliocentric distance. The dotted lines in Figures \ref{fig:kde2} (c) and (d) correspond to the isotropic ({velocity} or magnetic) energy distribution. While the {normalized velocity fluctuation} ratios show a dependence on the heliocentric distance $r$ (with a very low amplitude), the magnetic {fluctuation ratios do} not show a clear dependence. However, the magnetic fluctuations are much larger than their means, while the {velocity} fluctuations are small when they are compared with their means. The variance anisotropy ratios, both {velocity} and magnetic, do not exhibit a dependence with respect to the heliocentric distance. Moreover, for the velocity field most of the cases remain around 2, suggesting that the kinetic energy distribution is approximately isotropic on  MHD scales, and for the magnetic field most of the events reported here show high anisotropy ratios (i.e., $2 \leq A_b$). 

\subsection{The incompressible energy cascade rate}\label{sec:epsi}

To  compute the right-hand side of Eqs.~\eqref{law_iso}, \eqref{law_perp}, and \eqref{law_para}, we constructed temporal correlation functions of the different turbulent fields at different time lags $\tau$ in the interval [1,3600] s, which  covers the MHD inertial range \citep{H2017a} {at heliocentric distances between $\sim$0.2 and $\sim$0.8 au}. Once we have the energy cascade rates as a function of the time increments, we average them on the large timescales (i.e., for $\tau\in[1000,3000]$ s) to obtain representative values for the cascades in the largest MHD scales.

\begin{figure}
    \centering
    \includegraphics[width=.45\hsize]{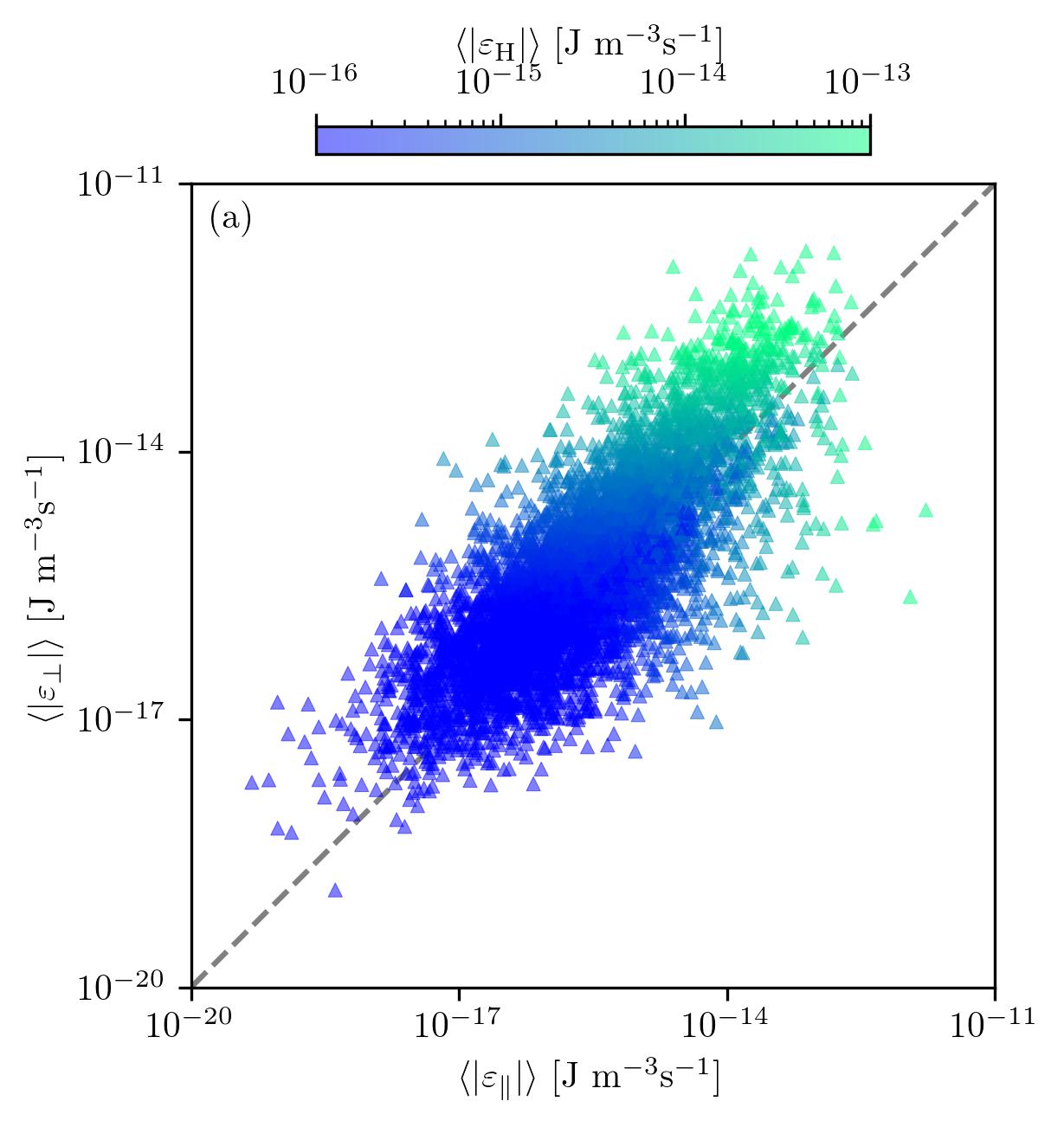}\includegraphics[width=.45\hsize]{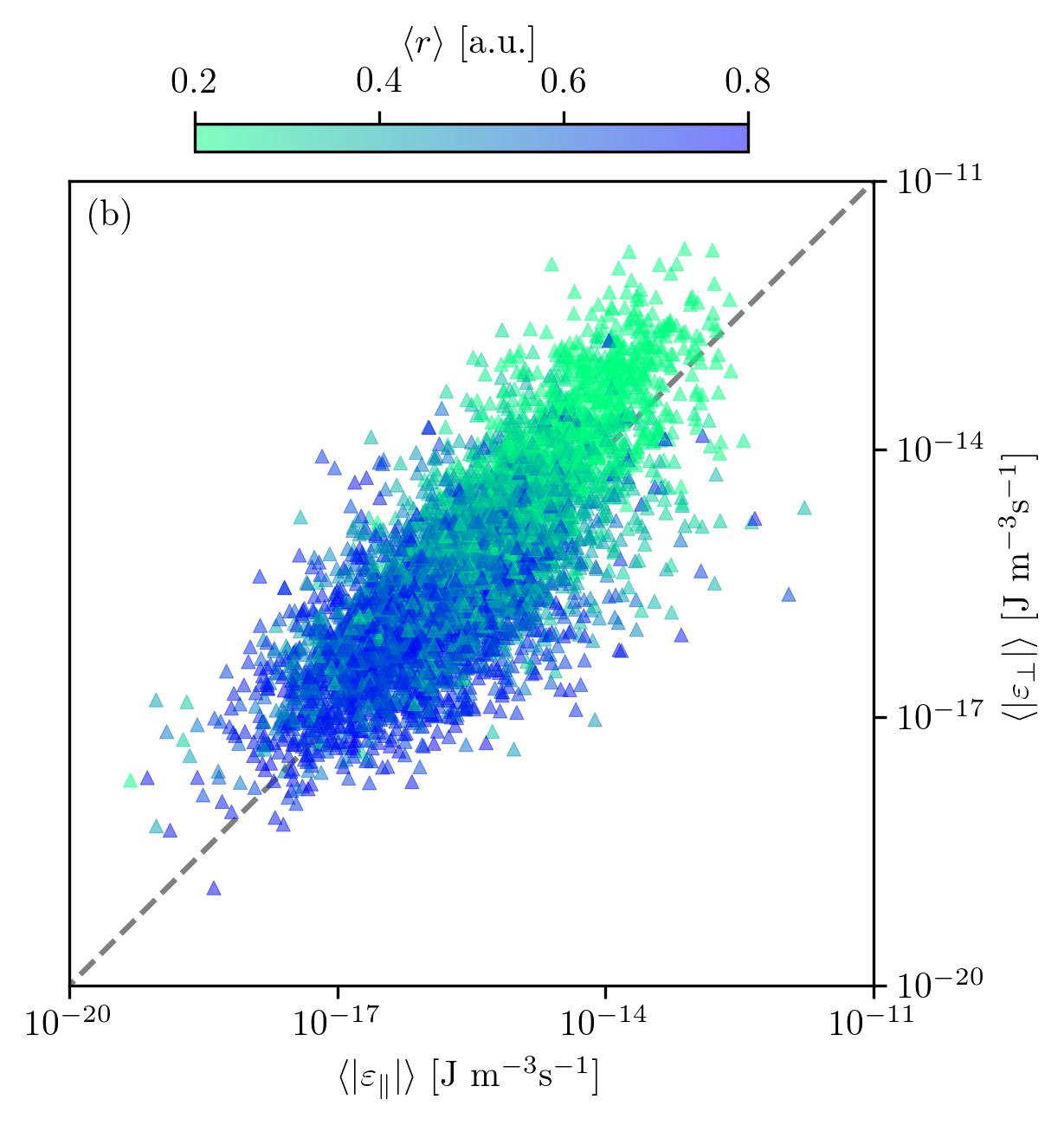}
    \caption{ Cascade rate component $\langle|\varepsilon_\perp|\rangle$ as a function of the component $\langle|\varepsilon_\parallel|\rangle$. In panel (a) the color bar is the total cascade $\langle|\varepsilon_\text{H}|\rangle$ {(where $\varepsilon_\text{H}=\varepsilon_\perp/2+\varepsilon_\parallel/4$ is the total hybrid cascade rate),} and in panel (b) it is  the heliocentric distance $\langle r \rangle$.}\label{fig:epsp}
\end{figure}

As we discuss in Section \ref{sec:iso}, the total isotropic energy cascade rate can be written as a function of two components,
\begin{align}
        \varepsilon_\text{I} &= \varepsilon_1+\varepsilon_2 ,\\ \label{eps1}
        \varepsilon_1 &= \rho_0\langle (\delta\uh\cdot\delta\uh+\delta\ua\cdot\delta\ua)\delta{u}_\ell /(-4 \tau U_0/3)\rangle, \\\label{eps2}
        \varepsilon_2 &= - \rho_0\langle (\delta\uh\cdot\delta\ua+\delta\ua\cdot\delta\uh)\delta{B}_{\ell}/(-4\tau U_0/3)\rangle,
\end{align}
where we can relate the first component $\varepsilon_1$ to the total energy (kinetic plus magnetic)  and the second component $\varepsilon_2$ to the cross-helicity (i.e., ${\bf u}\cdot{\bf B}$) in the plasma. This interpretation comes directly from Eqs.~\eqref{eps1} and \eqref{eps2}.

Figure \ref{fig:epsi} shows the mean absolute value $\langle |\varepsilon_2| \rangle$ as a function of $\langle |\varepsilon_1| \rangle$. The color bar corresponds in panel (a) to  the mean total energy cascade rate absolute value $\langle |\varepsilon_\text{I}| \rangle$ and in panel (b) to the heliocentric distance $r$. As a reference, we plot a gray dashed line with slope equal to 1. As we expected, there is a strong correlation between the cascade rate amplitude and the heliocentric distance to the Sun: the closer  PSP is to the Sun, the  stronger  the isotropic energy cascade rate is. In particular, the strongest cases correspond to {approximately} equal cross-helicity and energy components {(i.e., $\langle |\varepsilon_1| \rangle \approx \langle |\varepsilon_2| \rangle$)}.

Figure \ref{fig:epsp} shows the mean absolute value $\langle |\varepsilon_\perp| \rangle$  as a function of $\langle |\varepsilon_\parallel| \rangle$ in the same format as in Figure \ref{fig:epsi}. As in Figure \ref{fig:epsi}, as we move far away from the Sun, both components decrease their amplitudes. Moreover, we observe a clear trend of  obtaining more perpendicular than parallel energy cascade values as we approach  the Sun (slope larger than one in Figure \ref{fig:epsp} (b)).

\begin{figure}
    \centering
    \includegraphics[width=.49\hsize]{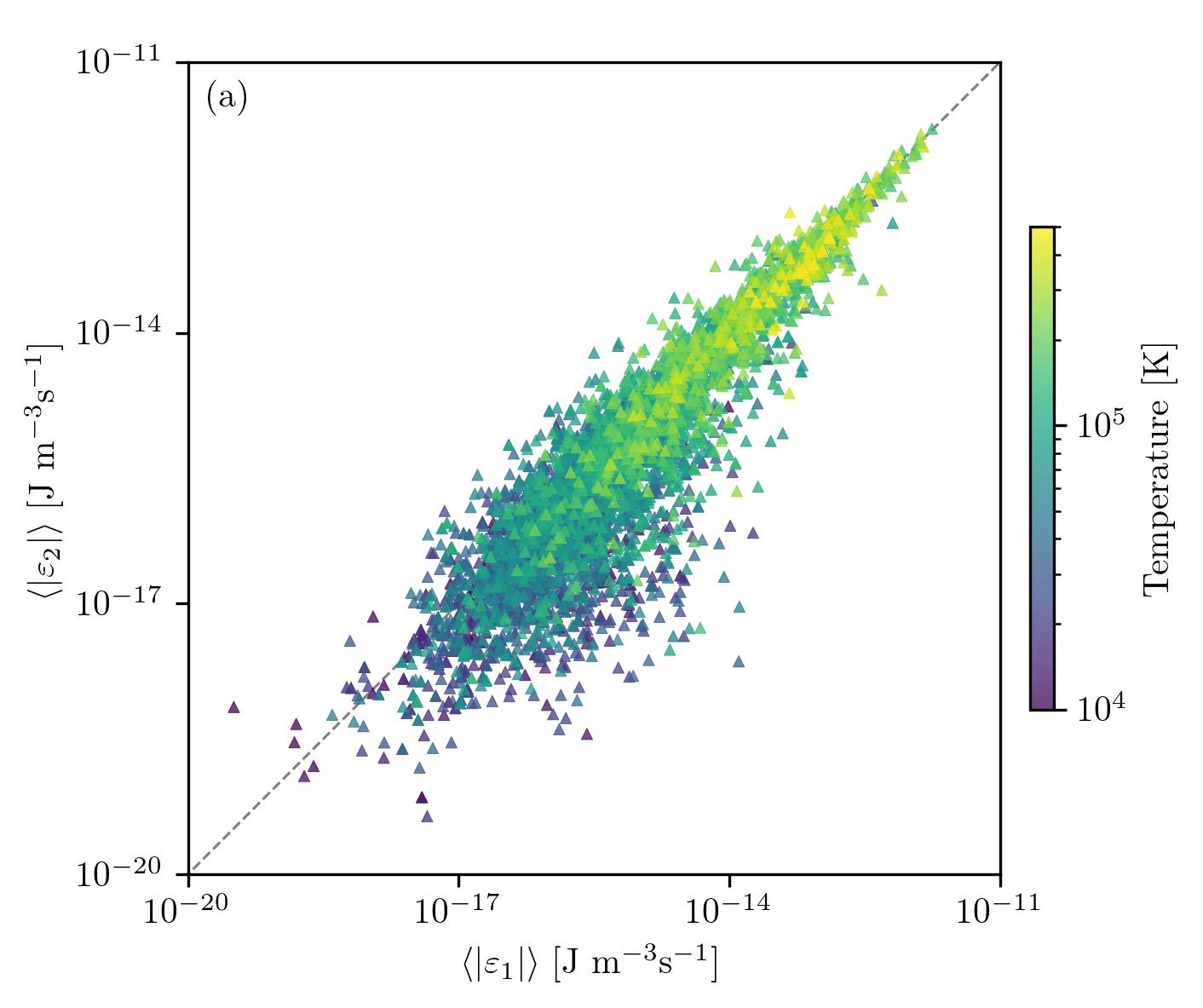}\includegraphics[width=.49\hsize]{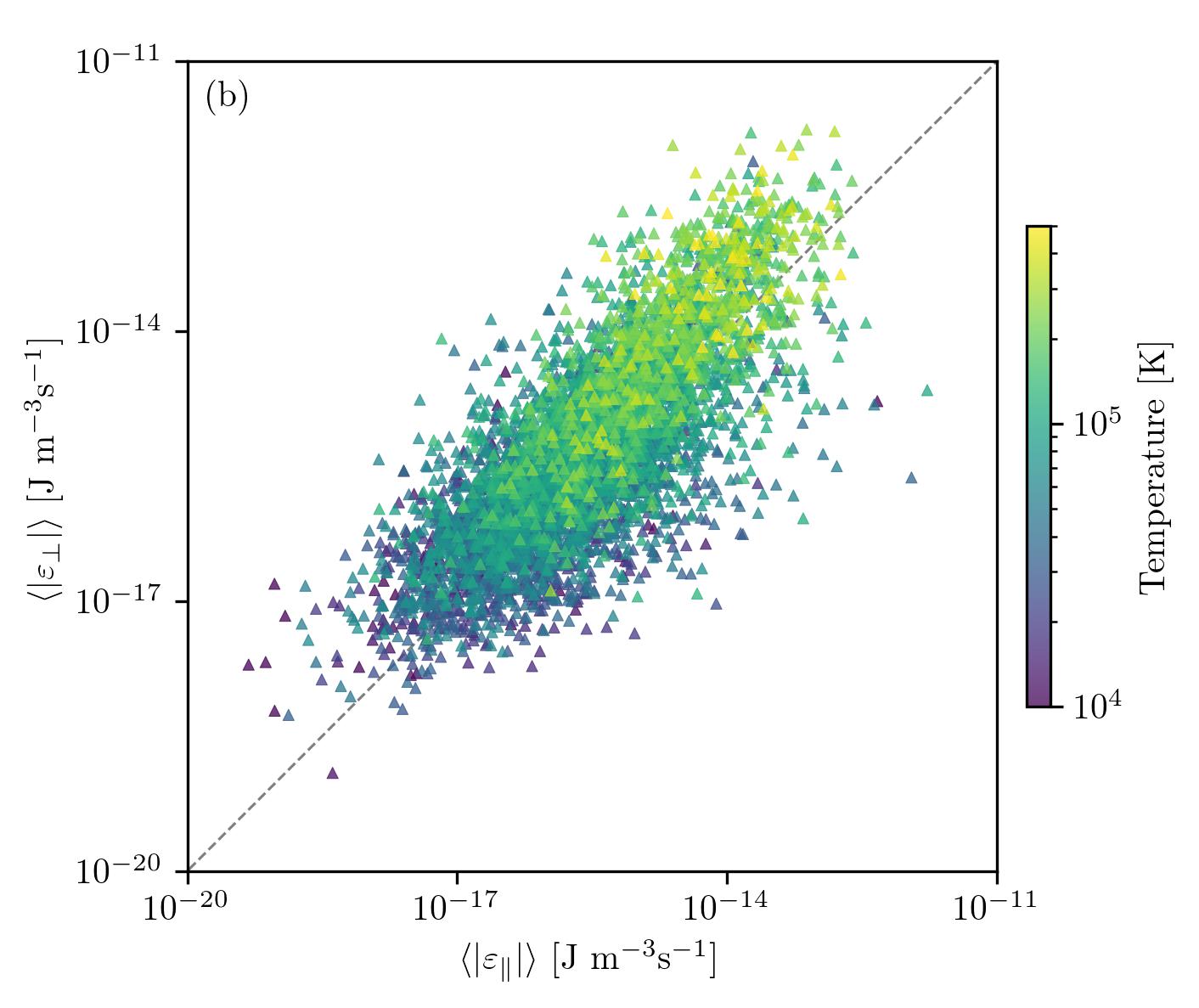}
    \caption{Cascade rate component $\langle\varepsilon_2\rangle$ as a function of the component $\langle\varepsilon_1\rangle,$ and the perpendicular component $\langle\varepsilon_\perp\rangle$ as a function of the parallel component $\langle\varepsilon_\parallel\rangle$. In  panels (a) and (b) the color bar is the temperature.}\label{fig:temp1}
\end{figure}

\begin{figure}
    \centering
    \includegraphics[width=.95\hsize]{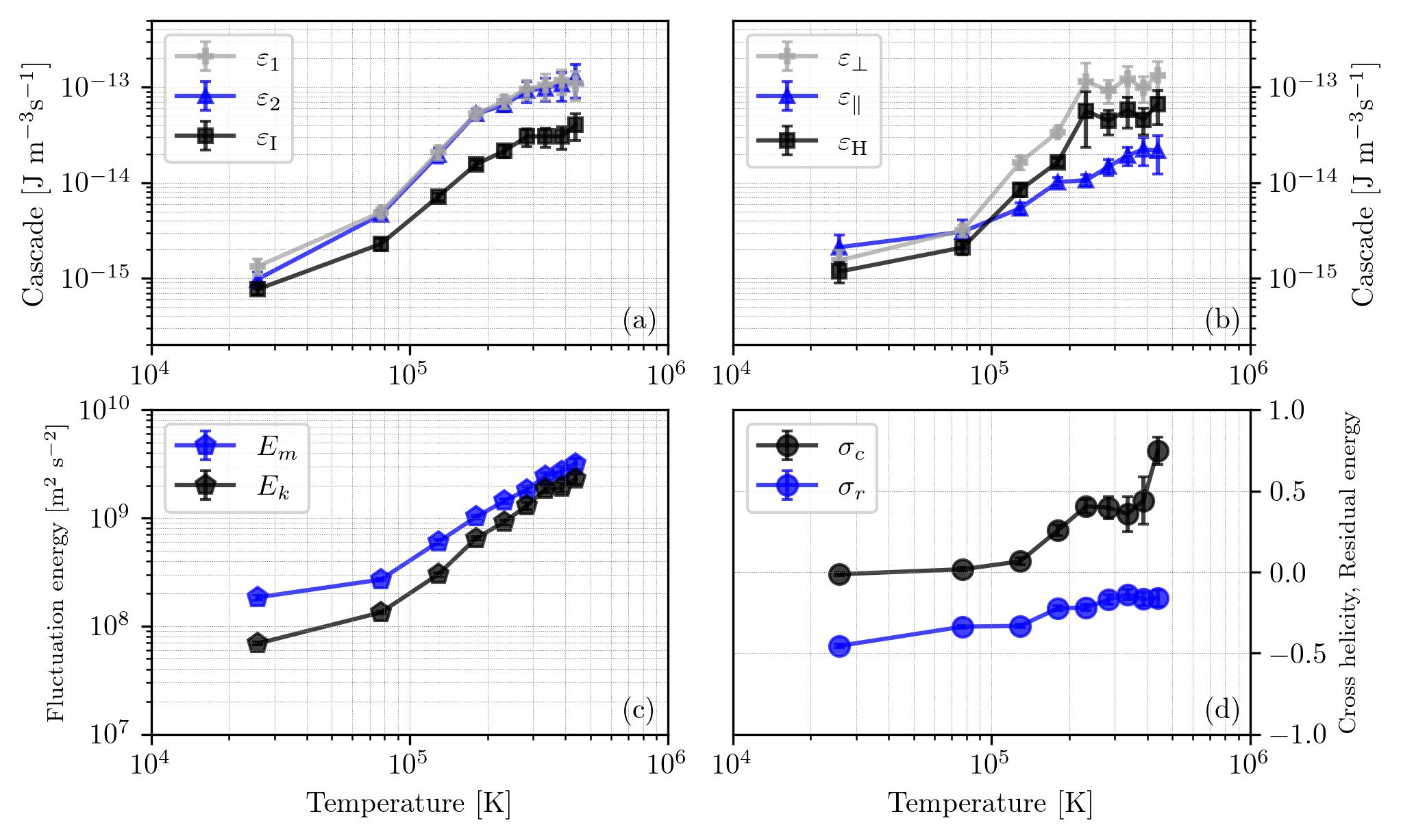}
    \caption{ \ADD{For a given temperature bin the following averages are shown:} (a) Components and total isotropic energy cascades rates; (b) components and total anisotropic energy cascades rates; (c) fluctuation in kinetic and magnetic energies; and (d) normalized cross-helicity and normalized residual energy as a function of the temperature.}\label{fig:temp2}
\end{figure}

\subsection{The isotropic, perpendicular, and parallel cascade rates and their relation with the temperature}

Figure \ref{fig:temp1} shows the mean absolute value $\langle |\varepsilon_2| \rangle$ as a function of $\langle |\varepsilon_1| \rangle$ and the mean absolute value $\langle |\varepsilon_\perp| \rangle$ as a function of $\langle |\varepsilon_\parallel| \rangle$. In both panels, the color bar corresponds to the proton temperature, and as a reference a gray dashed line indicates a  slope equal to one. In comparison with Figures \ref{fig:epsi} and \ref{fig:epsp}, we note the clear (and expected) correlation between the heliocentric distance and the temperature: as $r$ increases, the temperature decreases. In the case of the anisotropic cascade rates, we also observed that the hottest events mainly correspond to those where the perpendicular cascade is dominant with respect to the parallel cascade in the MHD range.

Typically, in MHD it defines the normalized cross-helicity $\sigma_c = \langle {\bf v}\cdot{\bf b}\rangle/(E_k+E_m)$ and the normalized residual energy $\sigma_r = (\langle {\bf v}^2 \rangle - \langle {\bf b}^2 \rangle)/(\langle {\bf v}^2 \rangle + \langle {\bf b}^2 \rangle)$, where $E_k\equiv\langle {\bf v}^2 \rangle/2$ is the incompressible kinetic energy and $E_m\equiv\langle {\bf b}^2 \rangle/2$ is the magnetic energy. While the cross-helicity measures the level of Alfv\'enicity of a particular event, the residual energy quantifies the relative energy in kinetic and magnetic fluctuations. By definition, the  parameters $\sigma_c$ and $\sigma_r$ range between -1 and 1. For simplicity we  drop the “normalized” prefix, assuming the understanding that these imply the normalized versions $\sigma_c$ and $\sigma_r$. Figure \ref{fig:temp2} shows the average of different variables as a function of the temperature. In particular, we group events according to the temperature values and then bin average them. The error bars correspond to the standard deviation divided by the square root of the number of samples in each group. Then, for a given temperature, we  averaged (a) the isotropy and (b) anisotropy energy cascade rates (total and components), (c) the incompressible kinetic and magnetic fluctuation energies, and (d) the cross-helicity and residual energy.

Figure \ref{fig:temp2} (a) and (b) show in a compact form the results analyzed in  Figure \ref{fig:temp1}: as the isotropic (or anisotropic) energy cascade rate increases, the temperature increases in the plasma. In particular, for the isotropic cascade the events with the highest temperatures correspond to {$\langle|\varepsilon_1|\rangle\approx\langle|\varepsilon_2|\rangle$}, while for the anisotropic cascade these events correspond to $\langle|\varepsilon_\perp|\rangle>\langle|\varepsilon_\parallel|\rangle$. Interestingly, in these hottest events the kinetic and magnetic fluctuation energies become approximately equal. Moreover, these events {seems to be} Alfv\'enic events since $\sigma_c\shortrightarrow1$.

\section{Discussion and conclusions}\label{sec:dis}

In this   paper we analyzed a large PSP solar wind data set of $\sim$ 5200 events, covering observations from October 2018 to December 2020. Our statistical results show a clear correlation between the incompressible energy cascade rate, heliocentric distance, and plasma temperature in the inner heliosphere. In particular, for both isotropic and anisotropic rates, as we decrease the heliocentric distance, the energy cascade rates increase by several orders of magnitude. We  covered heliocentric distance from $\sim0.8$ au  to $ \sim0.1$ au, obtaining energy cascade rates from $\sim1\times10^{-19}$ J m$^{-2}$s$^{-1}$  to $\sim1\times10^{-12}$ J m$^{-2}$s$^{-1}$. Recently, \citet{B2020} estimated the isotropic energy cascade rate for the first PSP perihelion. The authors found that $\varepsilon_\text{I}$ at $\sim$ 0.17 au is about 100 times higher than the average value at 1 au. In agreement with this finding and previous statistical results \citep[see][]{Mac2008,A2021}, we  found an amplification of $\varepsilon_\text{I}$ and $\varepsilon_\text{H}$ as we approach  the Sun. This amplification as we decrease the heliocentric distance is due to the increase in the {velocity} and magnetic fluctuation amplitudes (see Figure \ref{fig:kde1}) and the mean solar wind density value.  

In contrast with previous results \citep{O2015}, we do not observe a clear dependence of the spectral and variance anisotropy ratios on the heliocentric distance in the inner heliosphere. \citet{O2015} reported a review of solar wind anisotropy with different anisotropy ratios $A_v$ and $A_b$ from slow and fast solar wind at different heliocentric distances. \citet{B1999}  computed $A_v$ and $A_b$ for three events at 0.3, 0.7, and 0.9. The authors found that the magnetic fluctuation variance ratio slightly increases with heliocentric distance, while the {velocity} ratio remains constant. On the other hand, using Helios 1 observations from 0.3 au to 1 au, \citet{Mac2010} showed that the magnetic variance anisotropy scales with both proton beta and the amplitude of fluctuation power spectrum with no dependence on the heliocentric distance. In agreement with \citet{Mac2010}, our statistical results do not show any apparent increase in $A_b$ (or $A_v$) with respect to the heliocentric distance. Moreover, we observe that most of the cases exhibit $A_b>A_v$ \citep[see][]{B1999} {in agreement with previous observational \citep{O2015} and numerical \citep{Ou2016} results}.

Using the isotropic assumption \citep{P1998a,P1998b} and the slab and 2D assumption \citep{Mac2008}, we computed the {incompressible} energy cascade rate components from both models {using PSP solar wind observations.} For the isotropic model, in the cases near the Sun (i.e., the largest cascade values or  hottest events) the energy and cross-helicity components (see Eqs.~\eqref{1} and \eqref{2}) are approximately equal. On the contrary, for the anisotropic model, in the same events the dominant component is the perpendicular one. At 1 au, using ACE solar wind observations from 1998 to 2005, \citet{Mac2008} reported different cascade values for different types of solar wind. The authors found that  fast and slow solar winds both exhibit an active cascade rate over the inertial range, and that the energy flux in the parallel cascade is consistently smaller than in the perpendicular cascade. Beyond the fact that we are exploring different heliocentric distances at different correlation times {(an independent event lasts two days or tens of correlation lengths at 1 au for MacBride et al., while we consider an event to last one hour or approximately four correlation lengths),} we observed the same trend: for a large majority of the cases the perpendicular cascade is much larger than the parallel one. This statistical result is totally consistent with a dominant 2D cascade and/or geometry in slow solar wind turbulence on   MHD scales \citep[e.g.,][]{Sh1983,M1996,Da2005,W2012,O2013,A2017a,B2021,Z2021,B2021}. Moreover, the NI   MHD model \citep[e.g.,][]{Z1993,Z2021} predicts that the energy-containing range in the slow solar wind is a superposition of a majority quasi-2D component and a minority slab component. Using the NI model, PSP observations, and Solar Orbiter observations, \citet{Z2021} and \citet{Ad2021}  show that both the slow and fast solar winds are not typically aligned with the large-scale magnetic field, and therefore the quasi-2D fluctuations are visible to the PSP spacecraft, in agreement with our findings here.

We found a robust correlation between the temperature, the heliocentric distance, and the isotropic and anisotropic energy cascade rates: as we approach  the Sun, the temperature and cascade rates both increase. The temperature rise is clearly related to the most Alfvénic events ($\sigma_c\shortrightarrow1$) in an imbalanced and magnetic fluctuation dominant regime {($E_m>E_k$ or $\sigma_r<0$)}. Using a NI MHD model, \citet{Z2021} predicted arbitrary values of the (normalized) residual energy with a tendency to evolve toward negative values in magnetic energy dominated regimes. The authors also analyze PSP slow solar wind observations showing that the normalized residual energy becomes increasingly negative with increasing heliocentric distance (i.e., it becomes magnetic energy-dominated with distance). In the present paper we confirm these predictions, exploring not only the heliocentric distance dependence, but also the amplification of the cascade and the local temperature. While we do not observe that $\sigma_r$ becomes increasingly negative with increasing heliocentric distance, we do observe a constant and negative value for $\sigma_r$ as we approach  the Sun. In addition, these observations of $\sigma_c$ and $\sigma_r$ are consistent with the dominant 2D structures over the minority slab component {\citep{B2008,B2011,Ou2016}}.

Finally, some aspects of this work require improvement. On the one hand, we did not take into account possible compressibility under various closures \citep{S2021b,S2021a}, which may be relevant even in the usual incompressible solar wind \citep{B2016c,H2017a,A2017a,A2021}. On the other hand, we did not include the sub-ion scales energy cascade physics \citep{A2018,A2019b,H2018,F2021a}, which are closely related to the solar wind heating problem {\citep[e.g.,][]{Ma2020,M2021}}. These issues are planned for  upcoming works.

\section{Acknowledgements}
N.A. acknowledge financial support from CNRS/CONICET Laboratoire International Associé (LIA) MAGNETO. N.A. acknowledges financial support from the following grants: PICT 2018 1095 and UBACyT 20020190200035BA. We thank the NASA Parker Solar Probe SWEAP team led by J. Kasper and FIELDS team led by S. D. Bale for use of data. {N.A. thanks M. Brodiano for fruitful discussions about the data set.}

\bibliographystyle{aa}

\end{document}